\input harvmac
\sequentialequations

\Title{\baselineskip=14pt\vbox{\rightline{DAMTP R98/13}
\rightline{hep-th/9803064}}}
{\vbox{\centerline{Statistical entropy of charged two-dimensional}
       \vskip4pt\centerline{black holes}}}

\centerline{Edward Teo}
\bigskip
\centerline{\sl Department of Applied Mathematics and Theoretical 
Physics, University of Cambridge,}
\centerline{\sl Silver Street, Cambridge CB3 9EW, England}
\medskip\centerline{and}
\medskip
\centerline{\sl Department of Physics, National University of 
Singapore, Singapore 119260}

\vskip 1.2in
\centerline{\bf Abstract}
\medskip
\noindent
The statistical entropy of a five-dimensional black hole in Type II 
string theory was recently derived by showing that it is U-dual to 
the three-dimensional Ba\~nados--Teitelboim--Zanelli black hole, and 
using Carlip's method to count the microstates of the latter. This 
is valid even for the non-extremal case, unlike the derivation 
which relies on D-brane techniques. In this letter, I shall exploit 
the U-duality that exists between the five-dimensional black hole 
and the two-dimensional charged black hole of McGuigan, Nappi and 
Yost, to micro\-scopically compute the entropy of the latter. It is 
shown that this result agrees with previous calculations using 
thermodynamic arguments.

\Date{}
\pageno=1

\lref\MNY{M.D. McGuigan, C.R. Nappi and S.A. Yost, Nucl. Phys. 
B375 (1992) 421.}

\lref\NP{C.R. Nappi and A. Pasquinucci, Mod. Phys. Lett. A7 (1992) 
3337.}

\lref\Hyun{S. Hyun, {\it U-duality between three and higher dimensional 
black holes\/} (hep-th/9704005).}

\lref\SS{K. Sfetsos and K. Skenderis, {\it Microscopic derivation of 
the Bekenstein--Hawking entropy formula for non-extremal black holes\/} 
(hep-th/9711138).}

\lref\GP{G.W. Gibbons and M.J. Perry, Int. J. Mod. Phys. D1 
(1992) 335.}

\lref\SV{A. Strominger and C. Vafa, Phys. Lett. B379 (1996) 99.}

\lref\BTZ{M. Ba\~nados, C. Teitelboim and J. Zanelli, Phys. Rev. 
Lett. 69 (1992) 1849.}

\lref\HW{G.T. Horowitz and D.L. Welch, Phys. Rev. Lett. 71 (1993) 
328.}

\lref\HH{J.H. Horne and G.T. Horowitz, Nucl. Phys. B368 (1992) 444.}

\lref\Bekenstein{J.D. Bekenstein, Lett. Nuov. Cim. 4 (1972) 737; 
Phys. Rev. D7 (1973) 2333; Phys. Rev. D9 (1974) 3292.}

\lref\Hawking{S.W. Hawking, Nature 248 (1974) 30; 
Commun. Math. Phys. 43 (1975) 199.}

\lref\Carlip{S. Carlip, Phys. Rev. D51 (1995) 632; 
Phys. Rev. D55 (1997) 878.}

\lref\MS{J.M. Maldacena and A. Strominger, Phys. Rev. Lett. 77 
(1996) 428.}

\lref\JKM{C.V. Johnson, R.R. Khuri and R.C. Myers, Phys. Lett. 
B378 (1996) 78.}

\lref\BMPV{J.C. Breckenridge, R.C. Myers, A.W. Peet and C. Vafa, 
Phys. Lett. B391 (1997) 93.}

\lref\BLMPSV{J.C. Breckenridge, D.A. Lowe, R.C. Myers, A.W. Peet, 
A. Strominger and C. Vafa, Phys. Lett. B381 (1996) 423.}

\lref\CM{C.G. Callan and J.M. Maldacena, Nucl. Phys. B472 (1996) 
591.}

\lref\HS{G.T. Horowitz and A. Strominger, Phys. Rev. Lett. 77 
(1996) 2368.}

\lref\HMS{G.T. Horowitz, J.M. Maldacena and A. Strominger, Phys.
Lett. B383 (1996) 151.}

\lref\HLM{G.T. Horowitz, D.A. Lowe and J.M. Maldacena, Phys. Rev. 
Lett. 77 (1996) 430.}

\lref\Maldacena{J.M. Maldacena, {\it Black holes in string theory\/}
Ph.D. Thesis, Princeton University (hep-th/9607235).}

\lref\BPS{H.J. Boonstra, B. Peeters and K. Skenderis, Phys. Lett. 
B411 (1997) 59.}

\lref\Strominger{A. Strominger, {\it Black hole entropy from 
near-horizon microstates\/} (hep-th/9712251).}

\lref\BH{J.D. Brown and M. Henneaux, Commun. Math. Phys. 104 (1986) 
207.}

\lref\Witten{E. Witten, Phys. Rev. D44 (1991) 314.}

\lref\MSW{G. Mandal, A.M. Sengupta and S.R. Wadia, Mod. Phys. Lett. 
A6 (1991) 1685.}

\lref\Peet{A.W. Peet, {\it The Bekenstein formula and string theory
(N-brane theory)\/} (hep-th/9712253).}

\lref\BHO{E. Bergshoeff, C. Hull and T. Ort\'\i n, Nucl. Phys. B451
(1995) 547.}

\lref\WZWitten{E. Witten, Commun. Math. Phys. 92 (1984) 455.}

\lref\HKS{S. Hyun, Y. Kiem and H. Shin, {\it Infinite Lorentz boost 
along the M-theory circle and non-asymptotically flat solutions in 
supergravities\/} (hep-th/9712021).}

\lref\Tseytlin{A.A. Tseytlin, Mod. Phys. Lett. A11 (1996) 689.}

String theory (and its extended brane descendants) has reached a 
critical stage at which it is able to begin addressing fundamental 
issues in quantum gravity, such as the statistical origin of black 
hole entropy. This, in particular, has been an outstanding problem 
ever since Bekenstein \Bekenstein\ proposed in 1972 that the entropy 
of a black hole is proportional to its area, and Hawking \Hawking\ 
subsequently found the constant of proportionality to be $1/4G$ 
(in units where $\hbar=1$), $G$ being Newton's constant. But in 
1996, Strominger and Vafa \SV\ used D-brane techniques to derive 
the Bekenstein--Hawking area formula for an extremally charged 
five-dimensional black hole in Type II string theory. They did 
this by mapping it to an equivalent weakly coupled D-brane 
configuration, and enumerating the microstates using a certain 
conformal field theory.

D-brane counting has also been extended to near-extremal black 
holes \refs{\CM,\HS}, as well as rotating \refs{\BMPV,\BLMPSV} and 
four-dimensional ones \refs{\MS,\JKM,\HLM}. In all these cases, 
precise numerical agreement with the Bekenstein--Hawking formula 
was found. However, this method crucially relies on supersymmetry, 
to ensure that BPS states do not receive quantum corrections when 
moving to and from the corresponding D-brane picture. Thus, it 
can only be applied to extreme, or at most slightly non-extremal, 
black holes. (For reviews of these calculations, see 
\refs{\Maldacena,\Peet}. The latter reference also describes more 
recent progress using Matrix theory.)

In a separate development, Carlip \Carlip\ succeeded in 
microscopically computing the entropy of the three-dimensional 
constant-curvature black hole of Ba\~nados, Teitelboim and Zanelli 
\BTZ. Although gravity in three dimensions has no local dynamics, 
the black hole horizon induces an effective boundary to space-time 
which is dynamical. This boundary is described by a two-dimensional 
Wess--Zumino--Witten model \WZWitten, and Carlip derived the 
Bekenstein--Hawking formula by enumerating the states of the theory. 
This method, unlike the one using D-branes, does not assume 
supersymmetry and hence applies to the extreme and non-extremal 
cases alike. But it only appears to work in three dimensions.

Subsequently, Hyun \Hyun\ realised that the three-dimensional 
BTZ black hole is connected to the above-mentioned four- and 
five-dimensional charged black holes \refs{\HS,\HMS,\HLM} by 
U-duality. Thus, these black holes are equivalent as far as string 
theory is concerned, even though they look remarkably different as 
space-times. For instance, the BTZ black hole is asymptotically 
anti-de Sitter, while the higher-dimensional ones are asymptotically 
flat. Crucial in getting from one case to the other are the so-called 
shift transformations \BPS. These are T-duality transformations with 
respect to isometries that are space-like everywhere but 
asymptotically null. (This type of transformation is, in fact, not 
new. It was also found by Horowitz and Welch \HW, when they showed 
that the BTZ black hole is T-dual to a class of asymptotically flat 
three-dimensional black strings \HH.)

By exploiting this U-duality between the three- and higher-dimensional 
black holes, Sfetsos and Skenderis \SS\ have recently argued that 
Carlip's method in effect provides a way, if somewhat indirect, of 
microscopically computing the entropy of the four- and 
five-dimensional black holes. Unlike the D-brane counting in 
\refs{\CM,\HS,\HLM}, this derivation is valid even for arbitrarily 
non-extremal configurations. Furthermore, it suggests a general 
underlying WZW description of black hole entropy.

Hyun \refs{\Hyun,\HKS} has also pointed out that a similar U-duality 
exists between two-dimensional charged black holes and the four- and 
five-dimensional ones. The two-dimensional black holes in question 
were derived by McGuigan, Nappi and Yost \MNY, and in a different 
form by Gibbons and Perry \GP, from the heterotic string 
target-space action (with $\alpha^\prime=1$):
\eqn\action{{1\over16\pi G}\int{\rm d}^2x\,\sqrt{-g}\,
{\rm e}^{-2\phi}\left\{R+4(\nabla\phi)^2-\hbox{$1\over4$}F^2
+c\right\}.}
This action may also be regarded as the compactification of the 
Type II string action, with the gauge potential $A$ and effective 
central charge $c$ arising from the dimensional reduction of the 
Neveu--Schwarz two-form $B$. These black holes, like their 
higher-dimensional counterparts, have in general two horizons and 
are asymptotically flat. But because the Bekenstein--Hawking formula 
cannot be straightforwardly applied in two dimensions, a different 
method has to be used to calculate the entropy of these black holes. 
A thermodynamic approach using boundary terms of the action was 
adopted in \refs{\GP,\NP}.

In this letter, I shall explicitly establish the U-duality between 
the two-dimensional black hole and the five-dimensional one. It is 
then shown that the five-dimensional entropy obtained by the 
Bekenstein--Hawking formula agrees with that derived in \refs{\GP,\NP}. 
Since the five-dimensional black hole is in turn U-dual to the 
three-dimensional BTZ one, it follows that the two-dimensional 
and BTZ black holes are U-dual, and have identical entropies. By 
counting the microstates of the latter using Carlip's approach, we 
have in effect, a derivation of the statistical entropy for the 
two-dimensional black hole. This is also further evidence that 
higher-dimensional black holes may be described by much simpler 
lower-dimensional physics.

The ten-dimensional Type IIA supergravity solution that we begin 
with is by now a familiar one. Suppose the compact coordinates are 
$x_5,\dots,x_9$. Then it consists of a solitonic NS 5-brane wrapping 
around these five directions, a fundamental string wrapping around 
$x_5$, and a gravitational wave propagating along $x_5$. The metric 
in the string frame is \refs{\Tseytlin,\HMS,\Maldacena}
\eqnn\BHi
$$ \eqalignno{{\rm d}s^2&=-(H_1K)^{-1}f{\rm d}t^2+H_1^{-1}K
\left({\rm d}x_5-(K^\prime{}^{-1}-1){\rm d}t\right)^2
+H_5(f^{-1}{\rm d}r^2+r^2{\rm d}\Omega_3{}^2)
\cr&\quad+{\rm d}x_6{}^2+\cdots+{\rm d}x_9{}^2,&\BHi} $$
where $r^2=x_1{}^2+\cdots+x_4{}^2$, and ${\rm d}\Omega_3{}^2$ is 
the metric on the unit three-sphere. The dilaton $\phi$ and 
Neveu--Schwarz two-form $B$ are given by
\eqn\antisymm{\eqalign{{\rm e}^{-2\phi}&=H_1H_5^{-1},\qquad 
B_{05}=H_1^\prime{}^{-1}-1+\tanh\alpha\,,\cr 
&\qquad B_{056789}=H_5^\prime{}^{-1}-1+\tanh\beta\,,}}
where $B_{05}$ is the `electric' field of the fundamental 
string, and $B_{056789}$ is the `electric' field that is dual to 
the `magnetic' field of the 5-brane with components $B_{ij}$, 
$i,j=1,\dots,4$ \SS. In the preceding expressions,
\eqn\nameless{H_1=1+{r_0^2\sinh^2\alpha\over r^2}\,,\qquad
H_1^\prime{}^{-1}=1-{r_0^2\sinh\alpha\cosh\alpha\over r^2}H_1^{-1},}
define harmonic functions of $r$ that characterise the fundamental 
string, for some parameters $r_0$ and $\alpha$. The corresponding 
functions $H_5$, $H_5^\prime$ for the 5-brane, and $K$, $K^\prime$ 
for the wave are defined in the same way, but with $\alpha$ 
replaced by $\beta$ and $\gamma$ respectively. Also,
\eqn\nameless{f=1-{r_0^2\over r^2}\,.}
The extremal limit of this solution is obtained by taking 
$r_0\rightarrow0$ and $\alpha,\beta,\gamma\rightarrow\infty$, in 
such a way that $r_0\sinh\alpha$, $r_0\sinh\beta$ and 
$r_0\sinh\gamma$ remain finite.

Dimensionally reducing in the $x_5,\dots,x_9$ directions, and going 
to the Einstein frame, we have the metric
\eqn\fivedBH{{\rm d}s^2=-(H_1H_5K)^{-{2\over3}}f{\rm d}t^2
+(H_1H_5K)^{1\over3}(f^{-1}{\rm d}r^2+r^2{\rm d}\Omega_3{}^2)\,.}
This describes a five-dimensional black hole \refs{\Tseytlin,\HMS}, 
one of a class which has been extensively studied using D-branes 
\refs{\SV{--}\HS}. There is an outer horizon at $r=r_0$ and an 
inner one at $r=0$. Its entropy is given by the Bekenstein--Hawking 
formula:
\eqn\entropy{S={A_H\over4G_5}={2\pi^2r_0^3\cosh\alpha\cosh\beta
\cosh\gamma\over4G_5}\,,}
where $A_H$ is the area of the outer horizon and $G_d$ is Newton's 
constant in $d$ dimensions. This quantity is unchanged under 
U-duality \refs{\HMS,\Maldacena}, an important (and remarkable) 
fact that will be verified below.

To obtain the two-dimensional black hole from \BHi, we have to 
carry out a series of U-duality transformations. We first perform 
an $T_5ST_{6789}ST_5$ transformation, where $T_{ij\cdots}$ denotes 
T-duality along the $i,j,\dots$ directions, and $S$ is the S-duality 
of Type IIB string theory. (The relevant rules can be found in \BHO.) 
The overall effect of these transformations is to exchange $H_5$ 
and $K$, i.e., convert the 5-brane into a wave and {\it vice 
versa\/}. The resulting metric is
\eqnn\BHii
$$ \eqalignno{{\rm d}s^2&=-(H_1H_5)^{-1}f{\rm d}t^2+H_1^{-1}H_5
\left({\rm d}x_5-(H_5^\prime{}^{-1}-1+\tanh\beta){\rm d}t\right)^2
\cr&\quad+K(f^{-1}{\rm d}r^2+r^2{\rm d}\Omega_3{}^2)
+{\rm d}x_6{}^2+\cdots+{\rm d}x_9{}^2. &\BHii} $$
In particular, the field $B_{056789}$ in \antisymm\ now appears 
in the off-diagonal part of this metric. We then apply a shift 
transformation on the wave \refs{\Hyun,\BPS,\SS}:
\eqn\shift{t\rightarrow(\cosh\beta)t+\exp(-\beta)x_5\,,\qquad
x_5\rightarrow{x_5\over\cosh\beta}\,.}
This is an SL(2,R) coordinate transformation which is well-defined on 
the cylinder $(t,x_5)$, and it is an element of the O(2,2) T-duality
group. \BHii\ becomes
\eqnn\nameless
$$ \eqalignno{{\rm d}s^2&=-(H_1H_5)^{-1}f{\rm d}t^2
+H_1^{-1}H_5\left({\rm d}x_5-(H_5^{-1}-1){\rm d}t\right)^2
+K(f^{-1}{\rm d}r^2+r^2{\rm d}\Omega_3{}^2)\cr
&\quad+{\rm d}x_6{}^2+\cdots+{\rm d}x_9{}^2, &\nameless} $$
but with $H_5$ now given by
\eqn\Hfive{H_5={r_0^2\over r^2}\,.}
This shift has also turned the radius $R$ of the compact $x_5$ 
coordinate into $R\cosh\beta$ \SS. Lastly, we again apply the same 
set of $S$ and $T$ transformations in reverse to return to the 
original configuration. The final metric is \BHi\ but with the
new $H_5$ in \Hfive.

Hence, we have removed the additive constant in the 5-brane harmonic 
function $H_5$ by a series of U-duality transformations. The resulting 
space-time is no longer asymptotically flat, and this would enable 
us to make the link with lower-dimensional black holes. (In fact, to 
recover the BTZ black hole, one has to repeat the above procedure for 
$H_1$ \refs{\Hyun,\SS}.)

Now, dimensionally reducing in the $x_5,\dots,x_9$ directions and 
setting $\alpha=\gamma$, the metric and dilaton become
\eqn\BHiii{\eqalign{&{\rm d}s^2=-\left(1-{r_0^2\over r^2}\right)
\left(1+{r_0^2\sinh^2\alpha\over r^2}\right)^{-2}{\rm d}t^2+\left(
{r^2\over r_0^2}-1\right)^{-1}{\rm d}r^2+r_0^2{\rm d}\Omega_3{}^2,
\cr&{\rm e}^{-2\phi}={r^2\over r_0^2}+\sinh^2\alpha\,.}}
Although the shift transformation \shift\ has made this metric 
appear singular in the extremal limit, a sensible result is 
obtained by judiciously rescaling the coordinates at the same time 
\SS. The extremal limit of \BHiii\ can be written as \Hyun
\eqn\BHextreme{{\rm d}s^2=-
\left(1+{r_1^2\over r^2}\right)^{-2}{\rm d}t^2
+{r_5^2\over r^2}{\rm d}r^2+r_5^2{\rm d}\Omega_3{}^2,\qquad
{\rm e}^{-2\phi}={r^2+r_1^2\over r_5^2}\,,}
where $r_1$ and $r_5$ are the respective limits of $r_0\sinh\alpha$ 
and $r_0\sinh\beta$ appearing in \BHi\ (with $\alpha=\gamma$). The 
metric in either case is the product of an asymptotically flat 
two-dimensional geometry and a three-sphere with constant radius. 
These two parts are completely decoupled from each other. As we 
shall see, the former describes two-dimensional charged black holes.

The two-dimensional black hole solution of McGuigan, Nappi and Yost 
\MNY\ is
\eqn\nameless{\eqalign{&{\rm d}s^2=-(1-2m{\rm e}^{-Qx}
+q^2{\rm e}^{-2Qx}){\rm d}t^2+(1-2m{\rm e}^{-Qx}
+q^2{\rm e}^{-2Qx})^{-1}{\rm d}x^2,\cr
&{\rm e}^{-2(\phi-\phi_0)}={\rm e}^{Qx},\qquad 
A=\sqrt{2}Qq{\rm e}^{-Qx}{\rm d}t\,,}}
where $m$ and $q$ are constants related to the mass and electric
charge of the solution respectively, with $m>0$ and $m^2\geq q^2$. 
$Q$ is a positive constant determined by the effective central 
charge $c$ appearing in the action \action. We have also explicitly 
included the additive constant $\phi_0$ of the dilaton. By changing 
the spatial variable to $y={\rm e}^{-Qx}$, this solution becomes
\eqn\MNYBH{\eqalign{&{\rm d}s^2=-q^2(y-y_1)(y-y_2){\rm d}t^2
+{1\over q^2Q^2}{{\rm d}y^2\over y^2(y-y_1)(y-y_2)}\,,\cr
&{\rm e}^{-2(\phi-\phi_0)}=y^{-1},\qquad A=\sqrt{2}Qqy{\rm d}t\,.}}
The asymptotic region is at $y=0$, while the two horizons are at 
$y_{1,2}=q^{-2}(m\mp\sqrt{m^2-q^2})$. The two-dimensional part of
\BHiii\ can be cast into this form by setting
\eqn\nameless{Cy={r_0^2\over r^2+r_0^2\sinh^2\alpha}\,,\qquad 
r_0={2\over Q}\,,}
where $C=2\sqrt{m^2-q^2}$. In particular, $Cy_1=1/\cosh^2\alpha$.
Although the relationship between $y$ and $r$ is singular in the 
extremal limit when $m=q$, an analogous transformation can be found 
in that case which is well-behaved (see (21) below). The extra 
factor of $C$ that appears in the dilaton term has to be redefined 
into $\phi_0$:
\eqn\redef{{\rm e}^{-2\phi_0}\rightarrow C{\rm e}^{-2\phi_0}.}
It can also be checked that $A_0$ agrees with $B_{05}$ up to a 
constant factor. This is due to the arbitrariness inherent in 
compactifying $x_5$; one is free to rescale this coordinate, and 
it will have the effect of rescaling $B_{05}$ as well.

Now, Newton's constant in a lower dimension is related to that in a 
higher one by \refs{\Maldacena,\Peet}
\eqn\nameless{G_d={G_{d+n}\over V_n}\,,}
where $V_n$ is the volume of the $n$-dimensional compactification 
manifold. Thus, in going from five dimensions to two, $G_5$ has to 
be divided out by the volume of the three-sphere, $2\pi^2r_0^3$. 
Furthermore, as mentioned above, the shift transformation \shift\ 
increases the radius of the compact $x_5$ coordinate by a factor 
of $\cosh\beta$. Hence, we have
\eqn\nameless{G_2={G_5\over2\pi^2r_0^3\cosh\beta}\,.}
The five-dimensional Bekenstein--Hawking entropy \entropy\ is then 
$S=\cosh^2\alpha/4G_2$. But inspecting the starting action that was 
used in \MNY\ shows that Newton's constant has been set to 
$1/16\pi{\rm e}^{-2\phi_0}$. After redefining $\phi_0$ as in 
\redef, the entropy becomes
\eqn\NPentropy{S=4\pi{\rm e}^{-2\phi_0}(m+\sqrt{m^2-q^2})\,,}
in agreement with the result that was derived in \NP. 

In the extremal case, the appropriate change of variables which turns 
the two-dimensional part of \BHextreme\ into \MNYBH\ is
\eqn\nameless{y={y_1r_1^2\over r^2+r_1^2}\,,\qquad r_1^2={4\over 
Q^2}m\,, \qquad r_5 = {2\over Q}\,,}
where the horizon at $r=0$ corresponds to $y=y_1=y_2$. Now, the entropy 
of the five-dimensional black hole \fivedBH\ in the extremal limit is 
$S=2\pi^2r_1^2r_5/4G_5$. Substituting the corresponding two-dimensional 
quantities into it, in particular $G_5=2\pi^2r_5^3G_2$, we again find 
agreement with the formula \NPentropy\ derived in \NP.

It can also be checked, in a similar manner, that the five-dimensional
Bekenstein--Hawking entropy formula agrees with that derived by Gibbons 
and Perry \GP\ for the two-dimensional black holes considered in their
paper. They found a solution of the form
\eqn\nameless{\eqalign{&{\rm d}s^2=-{(m^2-q^2)\sinh^22\lambda x\over
\left(m+\sqrt{m^2-q^2}\cosh2\lambda x\right)^2}{\rm d}t^2+{\rm d}x^2,\cr
&{\rm e}^{-2(\phi-\phi_0)}={1\over2}\left({m\over\sqrt{m^2-q^2}}+\cosh2
\lambda x\right),\cr &A={\sqrt{2}q\over m+\sqrt{m^2-q^2}\cosh2\lambda 
x}{\rm d}t\,,}}
where the constant $\lambda$ is the analogue of $Q$ in \MNYBH. Unlike 
\MNYBH, this solution only covers the region exterior to the outer 
horizon at $x=0$. The appropriate change of variables linking the 
two-dimensional part of \BHiii\ and this solution is
\eqn\nameless{{r^2\over r_0^2}={1\over2}(\cosh2\lambda x+1)\,,\qquad
r_0={1\over\lambda}\,,\qquad
\sinh^2\alpha={1\over2}\left({m\over\sqrt{m^2-q^2}}-1\right).}
In this case, the five-dimensional entropy \entropy\ then becomes
\eqn\GPentropy{S=2\pi{\rm e}^{-2\phi_0}\left({m\over\sqrt{m^2-q^2}}
+1\right),}
where we have again used the fact that $G_2=1/16\pi{\rm e}^{-2\phi_0}$. 
This expression for the entropy is identical to the one derived in \GP. 
Note that \GPentropy\ is singular in the extremal limit $m=q$, unlike 
the corresponding quantity in \NPentropy. This is because the two 
solutions are related by a change of variables that is singular in 
this limit (essentially the factor $C$ introduced above that was 
absorbed into $\phi_0$, as in \redef). We also note that the charge 
parameter $q$ in the above solution may be taken to be zero, in which 
case we recover the well-known Witten black hole \refs{\Witten,\MSW}.

We have therefore established the relationship between the 
five-dimensional black hole \BHi\ and its two-dimensional counterpart. 
In particular, we have checked that the Bekenstein--Hawking entropy 
of the former agrees with that derived in the two-dimensional case 
using thermodynamic methods. By the results of Sfetsos and Skenderis 
\SS, it follows that the two-dimensional black hole is also U-dual to 
the BTZ one, and have identical entropies. (Of course, one could 
establish this U-duality directly, but we have chosen the more scenic
route here to illustrate the important link with the almost ubiquitous
five-dimensional black hole.) Hence, we may count the microstates of 
the two-dimensional black hole by equivalently enumerating those of 
the BTZ black hole. As mentioned earlier, this was carried out by 
Carlip \Carlip, who made use of the WZW theory that is induced on 
the black hole horizon.

It should be mentioned that Strominger \Strominger\ has 
recently proposed another way of microscopically deriving the 
Bekenstein--Hawking formula for the BTZ black hole. Unlike Carlip's 
method, he uses the fact that any three-dimensional geometry which 
is asymptotically locally anti-de Sitter is described by a conformal 
field theory \BH. This means there is an isomorphism between the 
boundary theories at infinity and the horizon, and either method 
could be used for our purposes.

\nref\Kaloper{N. Kaloper, Phys. Rev. D48 (1993) 2598.}

\nref\ILS{N. Ishibashi, M. Li and A.R. Steif, Phys. Rev. Lett. 67 
(1991) 3336.}

\nref\CT{M. Cveti\v c and A. Tseytlin, Phys. Lett. B366 (1996) 95.}

\nref\CL{M. Cveti\v c and F. Larsen, Phys. Rev. D56 (1997) 
4994; Nucl. Phys. B506 (1997) 107.}

\nref\Satoh{Y. Satoh, {\it Propagation of scalars in non-extremal 
black hole and black $p$-brane geometries\/} (hep-th/9801125).}

An important property of these lower-dimensional black holes is 
that they have explicitly known {\it world-sheet\/} conformal field 
theories associated to them. (This is to be contrasted with Carlip's 
{\it space-time\/} WZW model.) The BTZ black hole is described by a 
WZW model of the group SL(2,R), divided out by a discrete subgroup 
\refs{\HW,\Kaloper}. The two-dimensional one is described by a gauged 
WZW model of the group (SL(2,R)/U(1)) $\times$ U(1) \refs{\ILS,\HH}, 
where the first factor characterises the space-time geometry 
\Witten\ and the second factor its charge. The U-duality between 
these and the higher-dimensional black holes suggests that the latter 
are also governed by the same conformal field theories. (Hints of 
this have been found in the past; see, e.g., \refs{\Tseytlin,
\CT{--}\Satoh}.) If this is the case, then otherwise intractable 
problems may be addressed by moving down to the lower-dimensional 
picture. Ironically, what has been branded as unrealistic toy models 
by some pundits, may well turn out to play an important r\^ole in 
understanding more realistic black holes.

\listrefs
\bye